\documentclass[prb,showpacs,preprint]{revtex4}
\usepackage{graphicx}
\usepackage{amsfonts}
\usepackage{times}

\begin{document}

\title{Overcoming limitations of nanomechanical resonators with simultaneous resonances}

\author{N. Kacem$^{1}$}
\email[]{najib.kacem@femto-st.fr}
\author{S. Baguet$^4$}
\email[]{Sebastien.Baguet@insa-lyon.fr} 
\author{L. Duraffourg$^{2,3}$}
\author{G. Jourdan$^{2,3}$}
\author{R. Dufour$^4$}
\author{S. Hentz$^{2,3}$}
\email[]{sebastien.hentz@cea.fr} \affiliation{$^1$FEMTO-ST Institute - UMR 6174, 
Applied Mechanics Department, F-25000 Besan\c con, France\\$^2$Universit\'e Grenoble Alpes, F-38000 Grenoble, France\\$^3$CEA, LETI, Minatec Campus, F-38054 Grenoble, France
\\$^4$Universit\'e de Lyon, CNRS INSA Lyon, LaMCoS UMR 5259, F-69621 Villeurbanne France}

\begin{abstract}
Dynamic stabilization by simultaneous primary and superharmonic resonances for high order
nonlinearity cancellation is demonstrated with an electrostatically-actuated,
piezoresistively-transduced nanomechanical resonator. We prove experimentally how the combination of both the third-order nonlinearity cancellation and simultaneous resonances can be used to linearly drive a nanocantilever up to very large amplitudes compared to fundamental limits like pull-in occurrence, opening the way towards resonators with high frequency stability for high-performance sensing or time reference.
\end{abstract}

\pacs{85.85.+j, 05.45.-a, 46.40.Ff, 46.15.Ff}

\maketitle

Nanoelectromechanical systems (NEMS) are being developed for many applications such as
ultrasensitive mass \cite{176,160} and gas sensing \cite{huang}, subattonewton force detection at
millikelvin temperature \cite{mamin} or single-spin detection \cite{spin}.  Recently and due to the
early occurrence of their nonlinearities \cite{Kacem}, resonant nano-scale mechanical devices have
been used as platforms to explore fundamental questions in nonlinear classical mechanics
\cite{Lifshitz}.

Large amplitude vibrations are essential to the frequency stability of a resonator, and hence to
the performance of a resonant nanosensor. The relative frequency noise spectral density of a
resonator within its bandwidth is given by \cite{117}:
\begin{eqnarray}
S_f(\omega)=\left(\frac{1}{2Q}\right)^2\frac{S_{x}(\omega)}{P_0}
\end{eqnarray}
where $Q$ is the quality factor, $S_{x}$ is the displacement spectral density and $P_0$ the displacement carrier power, {\it
i.e.} the root mean square drive amplitude of the resonator $\frac{1}{2}A^2$.

The maximum amplitude is usually set by the onset of nonlinearity \cite{128}, either for stability
reasons \cite{123} or noise mixing in the carrier side bands \cite{125}. Some work has been carried
out in the past few years to overcome this limitation through hysteresis suppression by
nonlinearity cancellation \cite{128, 140, Kacem}. This is usually achieved by adding a negative
contribution to the third order (Duffing) nonlinear term by means of an electrostatic gate for instance.
Unfortunately, this technique is limited by higher order nonlinear terms which may be of importance
even for small amplitudes, \emph{i.e.} a fraction of the gap between the resonator and the
electrode (typically below $20\%$ of the actuation gap for electrostatically-actuated NEMS)
\cite{Kacem}. The amplitude response can become a multivalued function of the frequency and five
possible amplitudes for one given frequency is a clear signature of the physical significance of
the fifth-order nonlinearities \cite{157}. It has been shown how the use of simultaneous primary
(drive force at $\omega_n$, resonance at $\omega_n$) and superharmonic (drive force at $\omega_n/2$,
resonance at $\omega_n$ due to the second order nonlinear term) resonances can tune and dynamically
delay the onset of this multivalued, highly unstable behavior \cite{SR}.

In this paper, we demonstrate experimentally how the combination of these two techniques, namely
the nonlinearity cancellation and simultaneous resonances, can be used to stabilize and linearly
drive a nanomechanical resonator up to very large amplitudes compared to fundamental limits like
pull-in occurrence.

\begin{figure}[!htbp]
  \begin{center}
    \includegraphics[angle=270,width=0.9\textwidth]{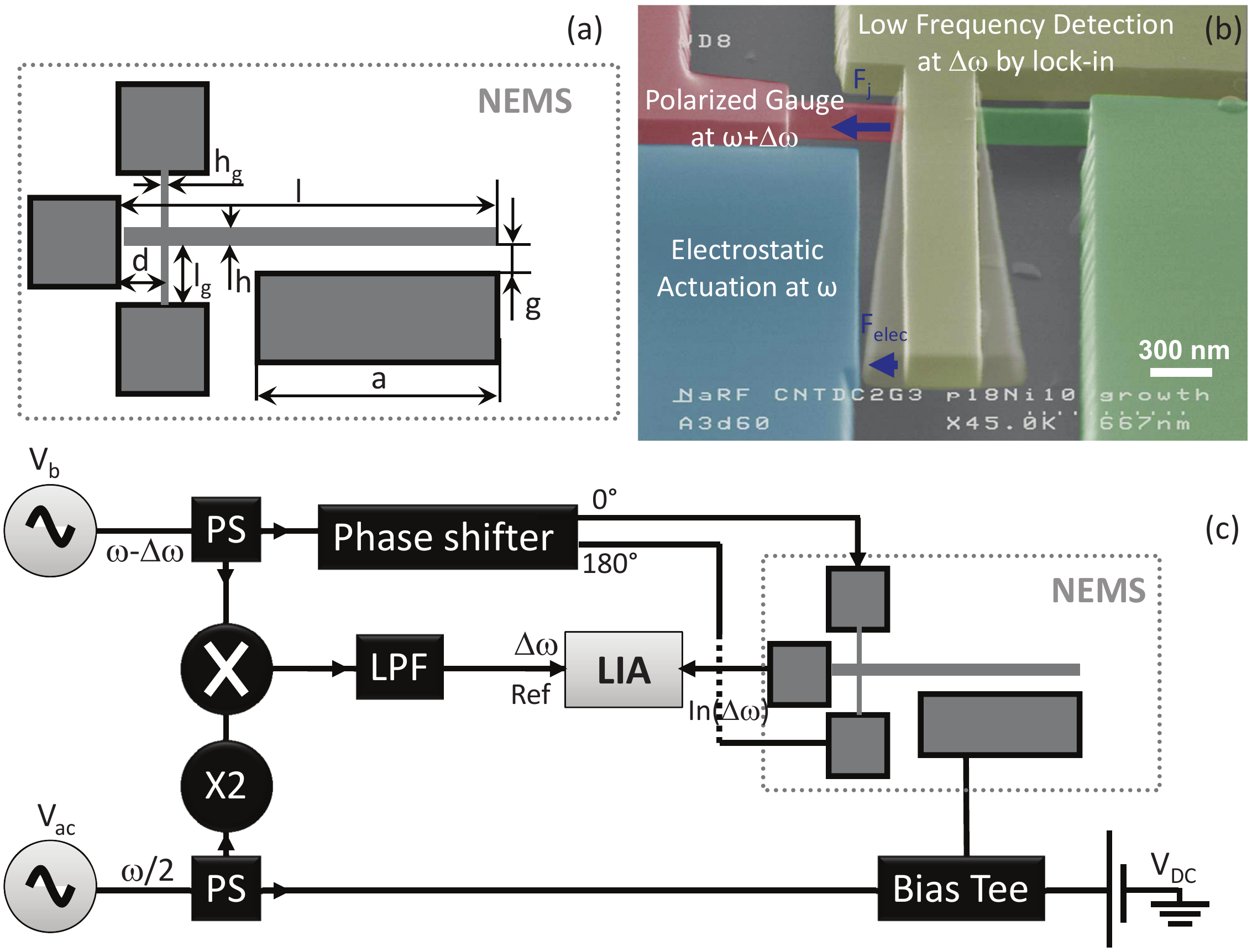}
  \end{center}
  \caption{(a): Design parameters of the NEMS device ($l=5\,\mu m$, $a=3.5\,\mu m$, $g=200\,nm$, $d=700\,nm$, $h=300\,nm$, $l_g=500\,nm$, $h_g= 80\,nm$).
  The gap $g$ is the in plane distance between the cantilever and the electrode. (b): Food-coloured SEM image of the in-plane piezoresistive structure.
  (c): Electrical measurement scheme of the piezoresistive NEMS device for both $2\Omega$ down mixing technique ($V_{dc}=0$) and simultaneous resonances ($V_{dc}\neq
  0$). $V_{ac}$ and $V_{dc}$ are respectively the alternating current ($AC$) and the  direct current ($DC$) voltages applied to the drive
  electrode. $V_{b}$ is the bias voltage applied to the nano-gauges, at a frequency slightly shifted from that of $V_{ac}$ to down-mix
  the output voltage at a low frequency (typically a few 10kHz).
  PS, LPF are power splitter and low pass filter, respectively.}
  \label{CB4}
\end{figure}

The NEMS device (see Figs. \ref{CB4}(a) and \ref {CB4}(b)) consists of a cantilever beam of length
$l=5\,\mu m$, electrostatically driven and connected to two suspended piezoresistive gauges at a
distance $d=0.15\,l$ from its anchored end. The cantilever vibration induces stress in the
piezoresistive gauges which in turn is transduced into a resistance variation. It is fabricated on
$200\,mm$ Silicon-On-Insulator wafers using complementary metal oxide semiconductor-compatible materials and processes. The readout
scheme and process is fully detailed in \cite{Mile}.

The considered NEMS has shown excellent performance in terms of frequency stability, with more than
$100\,dB$ dynamic range, and achievable mass resolution of few tens of zeptograms when operating in
the linear domain \cite{Mile, Hanay}. We describe here how the electrostatic actuation of this
nanoscale device is suitable to implement the simultaneous resonances and increase its dynamic
range even further.

A variational approach, based on the extended Hamilton principle \cite{156} has been used in order
to derive the nonlinear equations of motion describing the flexural vibration of the device
sketched in Figure \ref{CB4}(a). A reduced-order model is generated by modal decomposition
transforming the continuous problem into a multi-degree-of-freedom system consisting in ordinary
differential equations in time \cite{128, 140}. Assuming that the first mode is the dominant mode
and neglecting the nonlinear damping, the resonator dynamics can be described by the following dimensionless nonlinear ordinary differential equation for a single degree-of-freedom $a_1$:
\begin{eqnarray}
\ddot{a_1}+\mu\dot{a_1}+\omega_n a_1+\alpha_2a_1^2+\alpha_3a_1^3
+\alpha_5a_1^5\nonumber\\=4V_{ac}V_{dc}\zeta \cos\left(\Omega t\right)+V_{ac}^2\zeta \cos\left(2\Omega t\right) \label{E1}
\end{eqnarray}
The dot denotes derivation with respect to time, $\omega_n$ the natural frequency, $\alpha_2$,
$\alpha_3$, $\alpha_5$, the coefficients of the quadratic, cubic and quintic nonlinearities respectively \cite{Kacem}, $\mu$ the constant damping coefficient. The amplitudes of the
first and the second harmonic of the drive frequency $\Omega$ are respectively proportional to $V_{ac}V_{dc}$ and $V_{ac}^2$, where $V_{ac}$ and $V_{dc}$ are the $AC$ and $DC$ voltages applied to the nanoresonator and $\zeta$ is a constant that depends on the geometric parameters.

When $\Omega$ is tuned around $\frac{\omega_n}{2}$, the resonant response at $\omega_n$ is obtained
by both the first and the second harmonic $2\Omega$. More precisely, the $2\Omega$-excitation
mainly generates the primary resonance at $\omega_n$, while due to nonlinearities, the
$\Omega$-excitation actuates a superharmonic resonance at $\omega_n$. The response at $\omega_n$ is
thus made of simultaneous primary and superharmonic resonances. The latter one is generated via a
"slow" excitation compared to the resonant frequency. Experimentally, this can be achieved using
the "$2\Omega$ mode" readout scheme described in Fig. \ref{CB4}(c). By contrast, if the frequency
doubler is removed and the drive frequency is tuned around $\omega_n$ with $V_{dc}\neq
  0$ and $V_{ac}\ll V_{dc}$, an "$1\Omega$ mode" is simply
obtained for conventional primary resonance only.

The device under test was wire-bonded to a radio frequency (RF) circuit board and loaded in a
vacuum chamber for measurements at room temperature. In order to avoid signal shortage by parasitic
impedances, a down-mixing technique \cite{158} was used to readout the resistance variation of
the two nanowire gauges at a lower frequency $\Delta\omega$. Differential read-out was performed
with in-phase and out-of-phase bias voltages $V_{b}$ (one gauge is under compressive stress while
the second one is under tensile stress) \cite{Mile}. In order to analyze the dynamic behavior of the resonator under primary resonance and discuss the onset of nonlinearities, the conventional $1\Omega$ mode (frequency doubler removed) was first used. The cantilever was actuated using high $DC$ voltages and a fixed $AC$ bias voltage at the gauges ($V_{b}=1.56\,V$ peak-to-peak) so as to reach the nonlinear regime. The frequency response was measured using a lock-in amplifier in frequency sweep-up and sweep-down in order to obtain a full characterization of the resonator bifurcation topology. A constant quality factor $Q=5000$ was measured on linear frequency resonance curves.

\begin{figure}[!htbp]
  \begin{center}
    \includegraphics[width=0.9\textwidth]{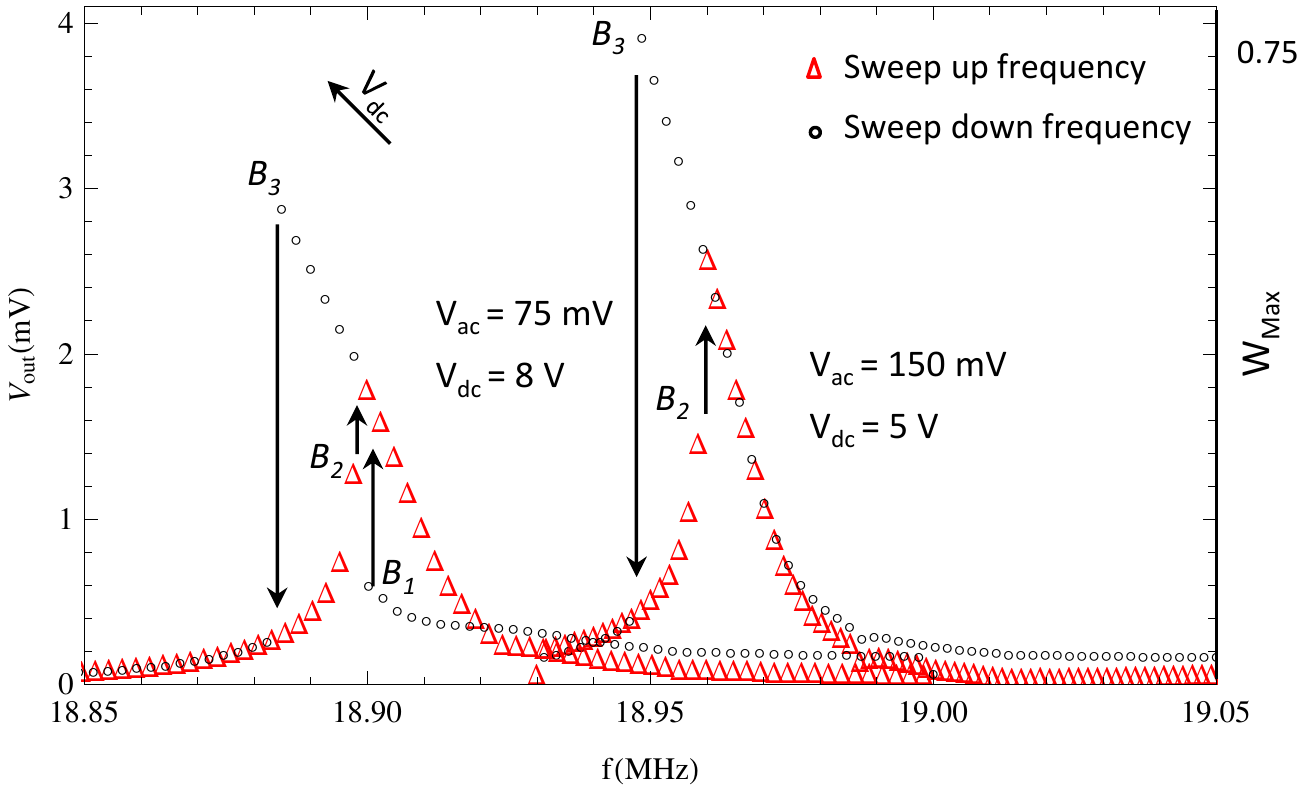}
  \end{center}
  \caption{Nonlinear resonance frequency responses measured using a $1\Omega$ down-mixing technique and showing the location of the different bifurcation points \{$B_1$, $B_2$ and $B_3$\}. $W_{max}$ is the cantilever displacement at its free end normalized by the gap. The hysteresis suppression is limited by high order nonlinear terms which leads to a highly unstable behavior for $V_{dc}=8\,V$ and $V_{ac}=75\,mV$.}
  \label{CB8}
\end{figure}

Figure \ref{CB8} shows two nonlinear resonance peaks in $1\Omega$ mode. The right-hand-side
resonance curve was obtained for $V_{ac}=150\,mV$ and $V_{dc}=5\,V $. It displays a hysteretic softening
behavior characterized by an amplitude jump-up at the bifurcation point $B_2$ and an amplitude jump-down at the bifurcation point $B_3$ where the cantilever oscillation amplitude is around
$75\%$ of the gap.

The left-hand-side resonance curve was obtained for $V_{ac}=75\,mV$ and $V_{dc}=8\,V $. Since the
actuation force is proportional to $V_{dc}.V_{ac}$, both voltages have a direct consequence on the
displacement amplitude: the resonance amplitude is lower than the one in the right-hand-side curve.
The increase in the $DC$ voltage also affects the resonator stiffness {\it i.e.} shifts down the
resonance frequency and amplifies the nonlinear electrostatic stiffness as well. Under those drive
conditions, the mechanical Duffing nonlinearity is negligible with respect to the electrostatic one
resulting in an amplified softening behavior (a stiffening behavior was also observed under different
experimental conditions). Despite a lower amplitude, the influence of the high-order nonlinear
terms and specifically of the quintic term ($\alpha_5 \propto V_{dc}^2$) is clearly visible: in frequency sweep-up, only one jump-up has been identified at the bifurcation point $B_2$. However, in frequency sweep-down, two jumps have been observed: a jump-up at the bifurcation point $B_1$ and a
jump-down at the highest bifurcation point in the softening domain $B_3$ where the cantilever
oscillation amplitude is around $52\,\%$ of the gap.

Obtaining a linear frequency response with this device at high amplitudes can be performed in
principle under primary resonance only ($1\Omega$ mode) by cancelling out the cubic term ($\alpha_3
(V_{dc})= 0$) in Equation (\ref{E1}) and thus suppressing the hysteresis. $\alpha_3$ depends
on the mechanical and electrostatic parameters and in particular on the $DC$ drive voltage.
Assuming that the second harmonic in Equation (\ref{E1}) is negligible, the optimal $DC$ drive
voltage cancelling $\alpha_3$ can be computed from \cite{Kacem}:

\begin{equation}
V_{dc_{OP}}=3\frac{\sqrt{\sqrt{5.16* g^{14} h^6+\frac{10^4g^{10} h^6 }{(l-d)^{-4}}}-4.53 g^7
h^3}}{10^{-9}(l-d)^4}\label{ee21}
\end{equation}
(The mechanical displacement between the anchors and the gauges being negligible, the device is
dynamically equivalent to a resonant nanocantilever of length $l-d$)

Applying Equation (\ref{ee21}) to the NEMS in use yields a relatively low optimal $DC$ voltage $V_{dc_{OP}}\approx1\,V$. Reaching large amplitudes (larger than the standard nonlinearity offset) in a
linear fashion would then require significantly high $AC$ voltages ($V_{ac}>0.5\,V$). Consequently,
the assumption of negligible second harmonic terms is not valid anymore and Equation (\ref{ee21}) is not applicable. More importantly, the non-linear terms of order five influence the dynamic behavior of the nanomechanical resonator already at low amplitudes as illustrated by the left-hand-side resonance curve in Figure \ref{CB8}. The nonlinearity cancellation is thus limited by potential dynamic instability such as pull-in phenomena. Finally, no linear and
stable behavior could be observed using the cancellation of the third-order terms with the device
under primary resonance only.

In order to overcome this issue, we combined the use of this third-order term cancellation
technique with the use of simultaneous primary and superharmonic resonances, which has been shown
to delay the onset of the quintic non-linear terms\cite{SR}. In this case, the
amplitude $X$ and phase $\beta$ modulations of the response can be written as:
\begin{eqnarray}
\dot{X}=-\frac{\epsilon\left(20 X\alpha _5 \zeta _1^2\cos\beta+9 \alpha _2 \omega _n^4\right)}{81 \omega _n^9\left(4 \zeta _1^2\sin\beta\right)^{-1}}-\frac{ \epsilon \mu  X}{2}+O(\varepsilon^2)\label{E7}\\
\dot{\beta}=-\frac{40 \epsilon  \alpha _5 \zeta _1^4\{6+\cos (2 \beta )\}}{81 \omega _n^9}-\frac{6 \epsilon  \alpha _3X^2 +5\epsilon  \alpha _5 X^4 }{16 \omega _n}\nonumber\\
+2\sigma-\frac{2\epsilon \zeta _1^2 \left(2 \alpha _2\cos  \beta +6 \alpha _3X +15\alpha _5 X^3 \right) }{9 \omega _n^5 X }+O(\varepsilon^2) \label{E8}
\end{eqnarray}
where $\varepsilon$ is the small nondimensional bookkeeping parameter and $\sigma$ is the detuning parameter.

In Equations (\ref{E7}) and (\ref{E8}), both slow (superharmonic) and fast (primary) dynamics are present due to terms proportional to $\alpha_5$ or $\alpha_2$, the coefficients of the quintic and quadratic nonlinearities. The $AC$ voltage sets the $2\Omega$ excitation while the $DC$ voltage only amplifies the $\Omega$ excitation. Unlike the case of primary resonance only shown in Figure \ref{CB8}, tuning the bifurcation topology in the present simultaneous resonances configuration can be obtained by simply increasing the $DC$ voltage, i.e. amplifying the superharmonic excitation amplitude.

\begin{figure}[!htbp]
\begin{center}
\includegraphics[width=0.9\textwidth]{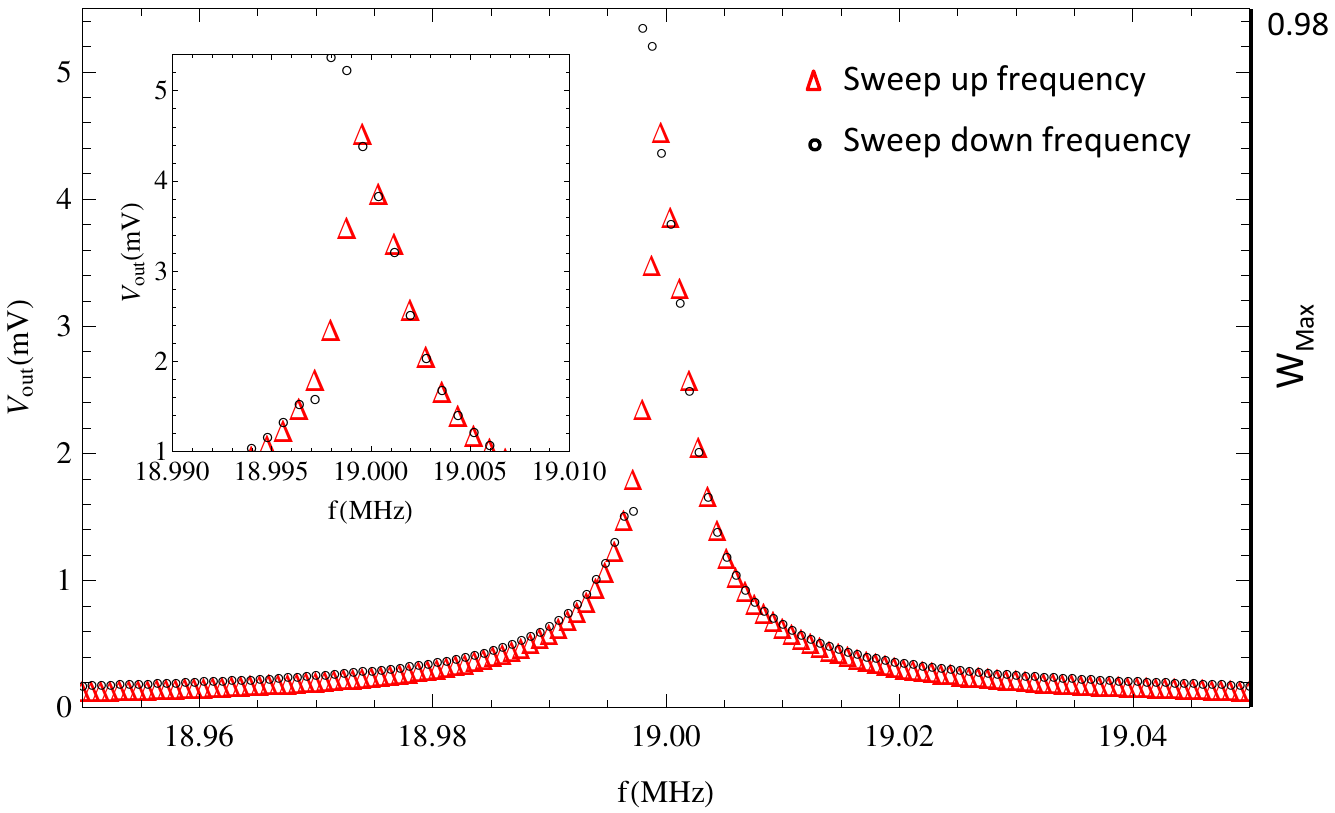}
\end{center}
\caption{(a): Slightly softening frequency response measured with the $2\Omega$ down-mixing
technique at the optimal $DC$ voltage $V_{dc}=0.5\,V$ for $V_{ac}=2\,V$ ($V_{b}=1.56\,V$). (b):
Zoom in the frequency region of the resonance peak. Unlike the peaks obtained under primary
excitation only in Figure \ref{CB8}, the combined use of third-order term cancellation with
simultaneous primary and superharmonic resonances enable a stable and linear behavior of the
nanomechanical resonator at very large amplitudes close to the actuation gap (the onset of
unstability being here above $95\,\%$ of the gap $g$).} \label{CB11}
\end{figure}

In practice, simultaneous primary and superharmonic resonances can be implemented extremely easily
with the $2\Omega$ downmixing readout scheme described in Fig \ref{CB4}(c), with $V_{dc}\neq 0$.
Moreover, high $AC$ voltages should be used to increase the first harmonic $\cos(\Omega t)$ in
Equation (\ref{E1}). The contribution of the $AC$ voltage in the nonlinear electrostatic stiffness
was taken into account in Equation (\ref{E1}) and the optimal drive $DC$ voltage was analytically and
numerically computed ($\alpha_3(V_{dc},V_{ac})=0$) for $V_{ac}=2V$, resulting in
$V_{dc_{OP}}\approx 0.5\,V$. Fig. \ref{CB11} shows a resonance peak obtained with these values. In
this case, the resonator behavior is almost perfectly linear up to extremely high amplitudes
compared to the actuation gap (above $95\%$ of the gap) and only a slightly softening behavior
remains. Computations predict that for this set of parameters the nonlinear electrostatic and
mechanical stiffnesses are balanced and that the oscillation amplitude of the cantilever is close
to $200\,nm$ at its free end (\emph{i.e.} the width of the gap). The
observed remaining softening part under these computed drive conditions are attributed to
fabrication uncertainties and approximations in $V_{dc_{OP}}$ calculations.

In this experiment, the second-order nonlinear term $\alpha_2$ giving rise to superharmonic
resonance is used to control the stability of the nanocantilever around its primary resonance by
retarding the effect of the fifth-order nonlinearity. The latter yields a highly unstable
behavior with five possible amplitudes for one given frequency at amplitudes below $50\%$ of the
gap $g$ as shown in Fig. \ref{CB8}. We demonstrate here how the use of simultaneous primary and
superharmonic resonances permits the dynamic stabilization of the nonlinear nanomechanical
resonator in order to maintain a linear behavior at large-amplitude vibrations close to the gap. At
this vibration level, the free end of the cantilever reaches a distance of a few nm to the
electrode without a damageable pull-in as would occur without the use of simultaneous resonance.

\begin{figure}[!htbp]
\begin{center}
\includegraphics[width=0.9\textwidth]{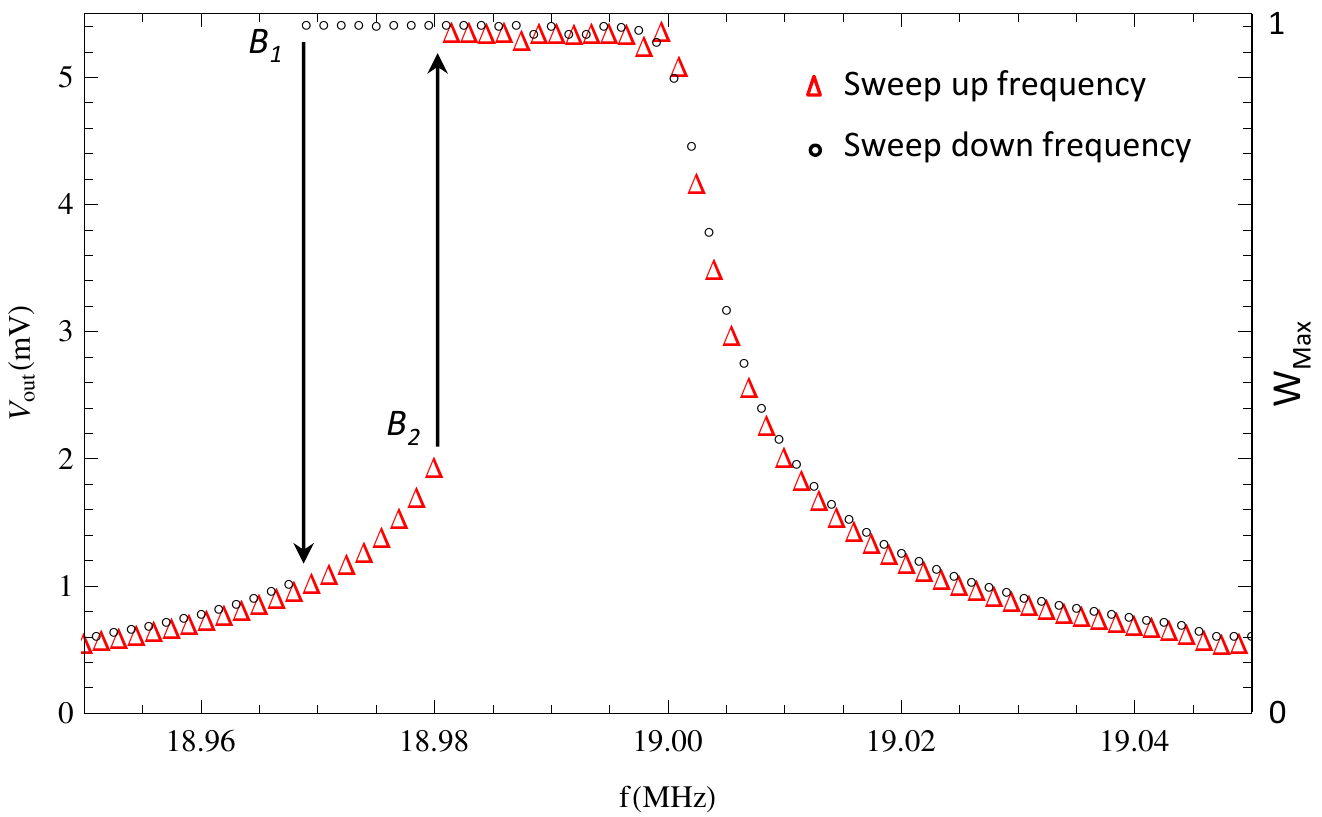}
\end{center}
\caption{Frequency response with $V_{dc}=V_{ac}=2\,V$ ($V_{b}=1.56\,V$). The horizontal branch
intercepting the bifurcation point $B_3$ is not due to any saturation in the readout scheme. It
rather shows that the cantilever free end comes in physical contact with the actuation electrode in
a reproducible way. This further confirms the high vibration amplitude reached by the device in
Fig. \ref{CB11}.} \label{CB12}
\end{figure}
To experimentally confirm such an oscillation level, the $DC$ voltage was gradually increased
 from $0.5\,V$ up to $2\,V$. Obviously, the vibration amplitude of the cantilever should increase, eventually hitting the actuation electrode.
 Besides, the cantilever behavior should subsequently become more softening. Figure
\ref{CB12} shows the frequency response obtained for $V_{dc}=2\,V$: the increase in the electrostatic
softening nonlinear stiffness is evidenced by the frequency shift between the two bifurcation
points (softening domain) which is significantly enlarged compared to Fig. \ref{CB11}. However the
output signal at the peak is still $V_{out} \simeq 5.4 mV$, \emph{i.e.} the same value as at
$V_{dc}=0.5\,V$. The observed plateau is not due to any kind of electrical saturation and is very
reproducible across multiple measurements. We attribute this to the fact that the resonator reached
a stable vibration amplitude equal to the gap (200nm). At one given bias voltage, this can be interpreted
as a measurement of the experimental output voltage sensitivity per unit of displacement of the
cantilever free end and confirms the very high amplitude reached in the measurement shown Fig.
\ref{CB11} compared to the gap.

In this letter, the nonlinearity cancellation has been demonstrated on a nanomechanical resonator
electrostatically actuated based on piezoresistive detection ($160\,nm$ thick) fabricated using a
hybrid e-beam/deep ultraviolet lithography technique. The device has been characterized using a $1\Omega$
down-mixing technique in its nonlinear regime to investigate the stability domain of the cantilever
dynamics under primary resonance. It has been shown that the high-order nonlinearities limit
drastically the operating domain of the nonlinearity cancellation. In order to overcome this limit,
the sensor has been actuated under its primary and superharmonic resonances simultaneously by using
a $2\Omega$ readout scheme. Specifically, the resonator can be linearly operated for
displacements well beyond the multistability limit and almost up to the gap, while retarding
undesirable behaviors and suppressing pull-in phenomenon.

This technique opens promising perspectives for time-reference or sensing purposes: the frequency
stability of the device should be significantly improved compared to its operation in linear regime 
below the bistability limit. In particular, the mass resolution one can expect with this device without reaching any damaging amplitude and
this technique is about $20$ zeptograms (five times smaller than the performance reported in
\cite{Mile}). Specific devices will be fabricated in the near future to substantiate this fact.

This work has been supported by the
CEA LETI and I@L Carnot institutes (NEMS Project). 

%
%

\end{document}